\documentstyle[12pt,aasms4]{article}

\begin{document}

\title{The Faint Optical Stellar Luminosity Function in the 
Ursa Minor Dwarf Spheroidal Galaxy }

\author{Sofia Feltzing\altaffilmark{1,2}}
\affil{Royal Greenwich Observatory, Madingley Road, Cambridge CB3 0EZ, UK}
\author{Gerard Gilmore}

\affil{Institute of Astronomy, Madingley Road, Cambridge CB3 0HA, UK}
\altaffiltext{1}{also 
Institute of Astronomy, Madingley Road, Cambridge CB3 0HA, 
UK}
\altaffiltext{2}{present address: Lund Observatory, Box 43, 221 00 Lund, Sweden}

\authoremail{sofia@astro.lu.se}
\authoremail{gil@ast.cam.ac.uk}
\author{Rosemary F.G. Wyse}
\affil{The Johns Hopkins University, Dept.~of Physics and Astronomy, 
Baltimore, MD 21218}
 \authoremail{wyse@pha.jhu.edu}
\begin{abstract}

Analyses of their internal stellar kinematics imply that the dwarf
spheroidal (dSph) companion galaxies to the Milky Way are among the
most dark-matter dominated systems known.  Should there be significant
dark matter in the form of faint stars in these systems, the stellar
luminosity function must be very different from that of a similar
metallicity globular cluster, for which there is no evidence for dark
matter.  We present the faint stellar luminosity function in the Ursa
Minor dSph, down to a luminosity corresponding to $\sim 0.45
M_{\odot}$, derived from new deep HST/WFPC2 data.  We find a
remarkable similarity between this luminosity function, and inferred
initial mass function, and that of the globular cluster M92, a cluster
of similar age and metallicity to the Ursa Minor dSph.

\end{abstract}


\section{Introduction}

Most of the dwarf spheroidal (dSph) companions of the Milky Way have stellar
velocity dispersions that are in excess of those expected if stars
with a normal mass function dominate their internal gravitational
potentials (see Mateo 1998 for a recent review).  Neither orbital
motions within stellar binaries (Hargreaves, Gilmore \& Annan 1996;
Olszewski, Pryor \& Armandroff 1996) nor Galactic tidal effects
(e.g. Piatek \& Pryor 1995) can inflate the internal velocity
dispersions enough to explain these measurements.  Rather, the stellar
kinematics imply the presence of gravitationally-dominant dark matter,
concentrated on small length scales, with 
mass-to-light ratio of a
factor of ten to fifty above those of normal stellar populations.
Could the dark matter be low-mass stars?  Stars of mass less than
about half that of the Sun have V-band mass-to-light ratio above 10 
(in solar units), and are
thus viable candidates, provided that the stellar initial mass
function (IMF) in these systems is very different from that of the
solar neighborhood, or in globular clusters.
  
A direct test of this hypothesis is provided by comparison of the
faint stellar luminosity function in a dSph galaxy with that of a
stellar system of similar age and metallicity but which is known not
to contain dark matter.  Empirical comparison between the luminosity
functions minimises the uncertainties in the transformation between
mass and light (see D'Antona 1998 for a recent discussion of this
point).  In addition to its possible relevance to dark matter
problems, the IMF of low-mass stars in a wide variety of astrophysical
systems is of considerable intrinsic interest (see e.g. papers in
Gilmore \& Howell 1998).
Various indirect means to determine the low-mass stellar IMF in
external galaxies have been applied, particularly analyses of optical and
infra-red surface brightnesses (Carr 1994).
The results are either inconclusive or apply only to the very outer
regions of galaxies (e.g. Davis, Feigelson \& Latham 1980; 
Gilmore \& Unavane 1998). Only direct star
counts can provide unambiguous results.

We here derive the faint stellar luminosity function in the Ursa Minor
dwarf spheroidal galaxy (distance modulus 19.11$\pm 0.1$, or $66 \pm
3$~kpc; Mateo 1998).  The internal stellar velocity dispersion of
$\sim 10$ km/s (Hargreaves et al.~1994; Olszewski, Aaronson \& Hill
1995) implies a total mass-to-light ratio of (M/L)$_V \sim 80$
(summarized in Mateo 1998).  This large mass-to-light ratio contrasts
with the typical value of (M/L)$_V$ for globular clusters of $\sim 2$
(e.g. Djorgovski \& Meylan 1994; Merritt, Meylan \& Mayor 1997).
Ground-based photometry of the red-giant branch and main-sequence
turnoff in the UMi dSph are consistent with essentially a mono-age
($\sim 15$~Gyr) and mono-metallicity ([Fe/H]$\sim -2.2$) stellar
population (Olszewski \& Aaronson 1985) similar to that of the old
metal-poor halo globular clusters, in particular to M92.  These narrow
ranges of age and metallicity contrast with most of the other dSph
companion galaxies in which there have been complex star-formation
histories and chemical evolution (e.g. Mateo 1998; Hernandez, Gilmore
\& Valls-Gabaud 1999).  The only other derivation of a luminosity
function for a dSph is that by Grillmair et al. (1998) for the upper
main sequence of the Draco dSph galaxy; in that case the analysis is
complicated by the significant metallicity spread in this galaxy
(e.g. Shetrone, Bolte \& Stetson 1998). Our analysis of the stellar
luminosity function in the Ursa Minor dSph is greatly simplified by
the lack of a significant spread in either age or metallicity, as this
lack implies a unique relationship between stellar luminosity and
mass. In this analysis we compare directly our HST luminosity function
for UMi with the luminosity function derived for M92 by Piotto, Cool
\& King (1997).

\section{Deep HST Star Counts in the Ursa Minor Dwarf Spheroidal Galaxy}

We obtained deep imaging data with the Hubble Space Telescope, using
all of WFPC2, STIS and NICMOS, in a field close to the center of the
Ursa Minor dSph (program GO~7419, PI~Wyse).  The correction of the
data in this field for contamination by foreground stars and
background galaxies required acquisition of similarly-exposed data for
a field at $\sim 2$ tidal radii away from the Ursa Minor dSph, but at
similar Galactic coordinates to the main UMi field ($\ell = 105^o, \,
b=45^o$).  The STIS and NICMOS data will be discussed elsewhere; here
we present the faint optical stellar luminosity function derived from
our WFPC2 data.

\subsection {Observations and Data Reduction}

The WFPC2 fields in the Ursa Minor dSph (hereafter `UMi') and offset
from the galaxy (`UMi-off') have WFALL coordinates ($\alpha_{\rm
2000}$, $\delta_{\rm 2000}$) = (15~08~00, +67~09~01) and (14~55~35,
+68~36~00) respectively.  Each field was exposed for $8 \times 1200$s
in each of F606W and F814W filters. Standard HST data reduction
techniques were followed, using the IRAF\footnote{IRAF is distributed
by National Optical Astronomy Observatories, operated by the
Association of Universities for Research in Astronomy, Inc., under
contract with the National Science Foundation.}  STSDAS
routines. Photometry on the reduced images used DAOPHOT, with the
recipe detailed by Cool \& King (1996), and TinyTim psf's, with
calibration following Holtzman et al.~(1995a,b). The scatter in the
zero points and photometric calibrations is $\sim 6\%$, providing a
calibration uncertainty which is small compared to the 0.5~magnitude
binning we adopt below.  Identical procedures were applied to both the
UMi and the offset datasets.

The main  feature of note in the present data reduction concerns the
completeness and reliability of faint source detection. 
Since we wish to go as faint as possible while
retaining reliability, we deliberately adopt a low value of the detection
threshold, and subsequently use two-color data to remove the inevitably
large number of spurious detections.  The appropriate detection
threshold was identified by running {\sc daofind} using different
values, and examining the number of detections as a function of $n
\times \sigma_{\rm bkgrnd}$. We adopted a threshold of $2.5\sigma_{\rm
bkgrnd}$, which provided 4000-5000 detections on each WF chip.  Two
statistics on the goodness of fit -- $\chi$ and sharpness -- are used
to determine which detections should be kept as bona fide stars and
which should be rejected. The sharpness parameter aids in the
separation between stars and resolved objects, such as background
galaxies.  An example of the ranges of the values of these parameters, 
for the WF2 image of the UMi field, is shown in
Figure~1, where the threshold values we adopted are 
indicated.  Again, the same parameter values were adopted for the
images of the offset field.  Finally, the stellar coordinates from the
independently-reduced F606W and F814W images were cross-identified,
using a matching radius of one pixel, to provide a list of all stellar
objects detected and measured in both V and I bands.

\subsection{Completeness Corrections}

The completeness of the data was determined by adding artificial stars
to the original images and then re-processing them as described above;
the fraction of the artificial stars that is recovered provides the
estimate of the completeness.  The number of artificial stars added
must be sufficiently high to provide reasonable statistical accuracy,
but not so high as to change the level of crowding in the image. In
each WF chip we detect in total about 450 stellar objects, so the
optimum number of artificial stars should be a few hundred for a given
magnitude.  We experimented with adding the stars either in a grid
pattern, with small random increments in $x$ and $y$ to avoid sitting
exactly at pixel centers, or completely randomly; full details are
described in Feltzing \& Gilmore (1999, where we also discuss tests
based on various subsets of the total integrations).

Briefly, 266 stars, all with the same magnitude, were added to the
F606W image of each WF chip. Each artificial star provides an
independent test.  The resulting images were analysed as described
above, giving the completeness for that particular magnitude.  This
process was repeated in 0.5~magnitude steps to the limiting magnitude
of the data.  The completeness function for the I-band data was
obtained through the same procedure, but with the magnitudes for F814W
chosen so that the mean ridge-line of the color-magnitude diagram
(CMD) was reproduced.  The completeness function for CMD-based
luminosity functions was obtained by the further requirement that the
recovered star should be detected in both V and I. The resulting
completeness function for the CMD-derived V-band luminosity function
(LF) is given in Table~1.  The estimated completeness functions for
the three WF chips agree extremely well, as detailed in Table~1,
reflecting the statistical robustness of our technique.  The
completeness function, based on the CMD, for the I-band LF is just
that for the V-band, but with the corresponding magnitude being offset by
the mean color of Figure 2(a).

The 50\% completeness of the 
V-band luminosity function derived on the basis of the CMD data 
is 27.25 mag, while that for the I-band is 50\% at 26.25 mag.

\section {The Color-Magnitude Diagrams}

Prior to comparison of the M92 and UMi luminosity functions, we first
confirm, by inspection of our derived CMD for UMi, that they both have similar
stellar populations and that our data appear reliable.  We consider
briefly the width of the main sequence, the presence of `blue stragglers', 
and the stars detected in the offset field.

The color-magnitude diagrams for the UMi and the UMi-off datasets are
shown in Figure~2(a,b), with 1-sigma photometric uncertainties as
indicated. Note that the ridge line of M92 follows that of our UMi data (not 
shown for clarity).
The apparent width of the main sequence in the UMi dSph is dominated
by photometric scatter. Any metallicity dispersion or line-of-sight
depth is undetected.  Unresolved binaries might affect the derived
luminosity function comparison, if the binary fraction or the binary
primary-secondary mass function were very different in the UMi dSph 
compared to that of M92.
We detect a marginally significant redwards asymmetry, of 0.02mag, in
the UMi color distribution, consistent with a small binary fraction.  For
the relevant mass range, the slope of the apparent luminosity function
is not sensitive to variations in binarism of this order (Kroupa, Tout
\& Gilmore 1991; their Figure 3), especially given the large bins in
apparent magnitude used below. The CMD of the UMi field contains a
small number ($\sim 5$) of stars blueward of the main-sequence
turnoff, consistent with the frequency of `blue
stragglers' derived by Olszewski \& Aaronson (1985). The CMD of the 
offset field contains no such stars, 
strengthening their identification with the 
Ursa Minor dSph. Interestingly, these stars are too blue to fit a
younger isochrone with the same metallicity as the dominant population
of UMi.  Their nature remains uncertain (more metal-poor?).  These few
stars are at brighter magnitudes than is relevant for our luminosity
function comparison.

The most obvious consequence of figure 2 is that the UMi-Off CMD
contains so few stars that contamination of the UMi CMD and luminosity
function by foreground stars or by background unresolved galaxies 
is not a concern. The star counts in the offset field are
also presented in Table~1, in the column UMi-Off.
Our luminosity function in the V-band derived from
the CMD is given in Table~1.  Note that we have confirmed that these
results are insensitive to the choice of bin center.

\section{Analysis and Discussion }

Care is needed in the choice of globular cluster data, since internal
dynamical relaxation effects and external tidal limitation modify the
observed faint luminosity (and mass) function from the initial one.
Mass segregation within a globular cluster depends on many unknown
parameters, including the initial binary fraction and the initial core
density.  Piotto, Cool \& King (1997) have determined the faint
stellar luminosity function for M92 in a field at $\sim 5$ times  
the cluster's half-light
radius.  Happily, a comparison of the luminosity functions derived
from two fields at 3.5 and 5 half-mass-radii within M92 shows that any
significant mass segregation is restricted to magnitudes fainter than
those of our comparison here (Andreuzzi et al.~1998).  Modifications
to the global LFs by external tidal effects such as disk shocking are
observed in many Galactic globular clusters.  Elson et al.~(1999) show
that the stellar mass function of M92 is essentially unmodified by
external dynamical effects. The Piotto et al. (1997) LF is therefore a
fair measure of the IMF in M92.
Their observations are in the same HST passbands
as our UMi data, and are reduced and calibrated in the same way as
adopted here. 

A straightforward comparison of the faint (unevolved stars) luminosity
functions in V and in I of the Ursa Minor dSph with those for M92
thus provides a direct comparison of the stellar IMFs 
in the range of interest here.  This is shown in
Figure~3(a,b) which contains the main result of this paper -- the
remarkable agreement between the faint stellar luminosity function of
a system in which there is no inferred dark matter and a system in
which very significant amounts of dark matter are inferred.

Further, these two systems -- a globular cluster and a dSph galaxy --
are at opposite extremes of stellar number density, with the central
V-band surface brightness of M92 being 15.6 mag/sq arcsec (Harris
1996), while that of the UMi dSph is 25.5 mag/sq arcsec (Mateo 1998).

The mass
corresponding to our 50\% completeness limit may be
estimated from stellar models.  The models of Baraffe et al.~(1997)
for low-mass stars of metallicity $-2$~dex, transformed to the
appropriate HST filters (their Figure~5) and at the distance and
reddening (E(V-I)=0.045) of UMi, translate these completeness levels
to masses of $\sim 0.45 M_\odot$.  The vandenBerg isochrones for
[Fe/H]$ = -2.2$~dex, transformed into the HST filters by Worthey
(priv.~comm.), provide consistent results. Should one wish to adopt a
mass--luminosity relation, then the resulting stellar mass function for the
UMi dSph would be essentially that derived by Piotto et al. (1997) for M92.

The observations presented here demonstrate that the faint stellar
luminosity function in the Ursa Minor dSph, measured down to a
luminosity corresponding to $\sim 0.45$M$_\odot$, is indistinguishable
from that of the globular cluster M92.  This precludes stars in this
mass range from being any part of the explanation of the high velocity
dispersions observed, and implied high dark matter content, in the UMi
dSph galaxy. Furthermore, the similarity of the stellar luminosity and
mass functions of the UMi dSph and of M92 over the whole range
observed to date provides no evidence for any difference in low-mass
IMF as a function of environment, for low metallicity stars.

\acknowledgements

Support for this work was provided by NASA through grant number
GO-7419 from the Space Telescope Science Institute, which is operated
by the Association of Universities for Research in Astronomy Inc,
under NASA contract NAS5-26555.  SF acknowledges financial support
from the Swedish Natural Research Council under their postdoctoral
fellowship program.  We thank Jay Gallagher, Rachel Johnson, Tammy
Smecker-Hane and Nial Tanvir for comments and discussions and are
grateful to Giampaolo Piotto for making the M92 data available to us.

\clearpage 
\figcaption{Statistical parameters $\chi$ and sharpness for 
WF2, F606W for the UMi dataset, versus V$_{606}$ magnitude. 
The applied cuts are shown as dashed lines.} 

\figcaption{The HST inflight color-magnitude diagram for (a) the UMi field 
and (b) the offset field.  The 
different symbols in (a) denote different WF chips. 
The 
1-sigma photometric uncertainties in color are as indicated; those in the 
magnitude are approximately a factor of 1.4 smaller and not shown for 
clarity.}

\figcaption{Comparison between the UMi completeness-corrected luminosity
functions derived from the CMD with that of the globular cluster M92
(filled squares); (a) the V-band and (b) the I-band.  The vertical
dashed lines indicate the 50\% completeness limits.  The
normalisations of the V- and I-band were applied using the CMD colors,
providing a consistency check on the data. }


\clearpage 

\begin{thebibliography}{}

\bibitem[]{} Andreuzzi, G., Buonanno, R., Iannicola, G. \& Marconi, G.
1998, Mem Soc Ast It, 69, 263


\bibitem[]{} Baraffe, I., Chabrier, G., Allard, F. \& Hauschildt, P. 1997, 
\aap, 327, 1054

\bibitem[]{} Carr, B. 1994, \araa, 32, 531

\bibitem[Cool and King 1996]{ck96} Cool A.M. \&  King I.R., 1996, 
HST postcalibration workshop, (STScI, Baltimore) p290

\bibitem[]{} D'Antona, F.  1998, in  The Stellar Initial Mass Function, 
ASP conf.~series 
vol 142, eds G.~Gilmore G. \& D.~Howell (ASP, San Francisco) p157


\bibitem[]{} Davis, M., Feigelson, E. \& Latham, D. 1980, \aj, 85, 131

\bibitem[]{} Djorgovski, S. \& Meylan, G. 1994, \aj, 108, 1292

\bibitem[]{} Elson, R.A.W, Tanvir, N., Gilmore, G., Johnson, R. 
\& Beaulieu, S. 1999, in New Views of 
the Magellanic Clouds, proc IAU Symposium 190,  eds Y.-H.~Chu, N.~Suntzeff, 
J.~Hesser \& D.~Bohlender, in press

\bibitem[]{} Feltzing, S. \& Gilmore, G. 1999, preprint



\bibitem[]{} Gilmore, G. \& Unavane, M. 1998, \mnras, 301, 813

\bibitem[]{} Gilmore, G. \& Howell, D. 1998, eds The Stellar IMF, 
ASP Conf Series 142 (ASP, San Francisco)

\bibitem[]{} Grillmair, C., Faber, S., Lauer, T. et al. 1998, \aj, 115, 144


\bibitem[]{} Hargreaves, J., Gilmore, G. \&  Annan, C. 1996, \mnras, 279, 108

\bibitem[]{}Hargreaves, J., Gilmore, G., Irwin, M.J. \& Carter, D. 1994, 
\mnras, 271, 693


\bibitem[]{} Harris, W.E. 1996, \aj, 112, 1487
\bibitem[]{} Hernandez, X., Gilmore, G. \& Valls-Gabaud, D. 1999,
\mnras, in press

\bibitem[]{} Holtzman, J., Hester, J., Casertano, S. et al. 1995a, \pasp, 107, 156

\bibitem[]{} Holtzman, J., Burrows, J.B., Casertano, S., Hester, J.J., 
Trauger, 
J.T., Watson, A.M. \&  Worthey, G. 1995b, \pasp, 107, 1065 










\bibitem[]{} Kroupa, P., Tout, C., \& Gilmore, G. 1991 MNRAS something


\bibitem[]{} Mateo, M. 1998, \araa, 36, 435


\bibitem[]{} Merritt, D., Meylan, G. \& Mayor, M. 1997, \aj, 1074


\bibitem[]{} Olszewski, E.W. \& Aaronson, M. 1985, \aj, 90, 2221

\bibitem[]{} Olszewski, E.W., Aaronson, M. \& Hill, J.M. 1995, \aj, 110, 2120

\bibitem[]{} Olszewski, E.W., Pryor, C. \& Armandroff, T. 1996, \aj, 111, 750

\bibitem[]{} Piatek, P. \& Pryor, C. 1995, \aj, 109, 1071

\bibitem[]{} Piotto, G., Cool, A. \& King, I.R. 1997, \aj, 113, 1345




\bibitem[]{} Shetrone, M., Bolte, M. \& Stetson, P. 1998, \aj, 115, 1888



\end{thebibliography}
\end{document}